\documentclass[a4paper]{article}
\usepackage[english]{babel}
\usepackage{epsfig}
\usepackage{amssymb}
\usepackage{amsmath}%
\usepackage{amsfonts}%
\usepackage{graphicx}

\begin{document}

\title{Ultrashort opposite directed pulses dynamics}
\author{Mateusz Kuszner, Sergey Leble\\\vspace{6pt} Gdansk University of Technology,
\\{\small ul. Narutowicza 11/12, 80-223,
 Gdansk, Poland,}}
\maketitle
\begin{abstract}
In this article we consider one dimensional model of an ultra short pulse propagation in isotropic dispersionless media taking into account a nonlinearity of the third order. We introduce a method for Maxwell's equations transformation based on a complete set of projecting operators. The operators generally correspond wave dispersion branches. As a simplest result of the method application we derive a system of equations describing dynamics of ultrashort pulses of opposite directions of propagation. We show that in such way the generalized Short Pulse Equations of Shafer and Wayne is obtained if the only directed wave is initialized. The effects of the unidirectional pulses interaction are traced.
\end{abstract}

\renewcommand{\abstractname}{\small }
\newcommand{\ed}{\textrm{d}}

\section{Introduction.}

Recently the trend towards shorter pulses duration account was successfully realized in a waverange $680 nm$ to infrared \cite{SchWayne}. The resulting equation (named as Short Pulse Equation - SPE) was properly investigated numerically \cite{ChungSchafer} and it was proven to be  integrable \cite{Sakovich, SakovichSakovich}. The derivation of the SPE was based on few approximations, included the dispersion relation adjusted to the mentioned waverange, third order nonlinearity account and unidirectionality of a pulse propagation realized via slow time variable introduction in the frame following the pulse.

A significant interest to the both directions of propagation account arises in different context, e.g. in resonator theory. One of examples of such account based on the second order wave equation is given in Ref. \cite{Kins1}. Author shows that unidirectional approach for hybrid electromagnetic fields allows to write down first order wave equation for pulse propagation. Moreover he shows in his article a generalization of undirectional \cite{Kins4} and bidirectional \cite{Kins3} approach presented in previous works.\\

Based on the interest and its wide development we suggest a general method \cite{
book:Leble0,book:Leble} implementation adjusted to the problems to be considered and which links unidirectional pulse propagation approach \cite{Kins2} with derivation of SPE \cite{SchWayne} and its generalizations.

In Sec. II we outline an idea and realization of projection operators that specify evolution operator subspaces. Later on we present an application of this approach to the derivation of bidirectional wave system for hybrid fields \cite{book:Leble,Kins4} and in final section our results  show in what conditions a transition to equations of earlier works \cite{SchWayne} are realized.

\section{Basic theory}
\subsection{The model outline}
Our starting point is the Maxwell equations for non-magnetic medium in the Lorentz-Heaviside's unit system
\begin{eqnarray}\label{eq_Maxwell}
\nabla\cdot\mathbf{D} &=& 0,\\
\nabla\cdot\mathbf{B} &=& 0,\\
\nabla\times\mathbf{E} &=& -\frac{1}{c}\frac{\partial\mathbf{B}}{\partial t},\\
\nabla\times\mathbf{H} &=& \frac{1}{c}\frac{\partial\mathbf{D}}{\partial t}
\end{eqnarray}
completed by material relations within the choice
\begin{equation}\label{eq_def_D}
\mathbf{D} = \mathbf{E} + 4\pi\mathbf{P}, \quad \mathbf{H} = \mathbf{B}.
\end{equation}
The polarization vector $\mathbf{P}$ have the form
\begin{equation}\label{eq_polariz}
\mathbf{P} = \mathbf{P}_{L}+\mathbf{P_{NL}}.
\end{equation}
We restrict ourselves by one-dimensional model as one of Ref. \cite{SchWayne},  x-axis chosen as a direction of a pulse propagation. We assume that $D_x = 0$ and $B_{x} = 0$ taking into account the only polarization of electromagnetic wave. This allows us to rewrite the Maxwell equations as
\begin{eqnarray}\label{eq_new_Maxwell's}
&&\frac{\partial}{\partial y}D_y + \frac{\partial}{\partial z}D_z =0,\quad \frac{\partial}{\partial y}B_y + \frac{\partial}{\partial z}B_z=0,\\\nonumber
&&\frac{1}{c}\frac{\partial D_y}{\partial t} = \frac{\partial B_z}{\partial x}\quad\textrm{where } D_{y} = E_{y} + 4\pi\chi^{(1)}E_{y} + (P_{NL})_{y},\\\nonumber
&&\frac{1}{c}\frac{\partial B_z}{\partial t} = \frac{\partial E_y}{\partial x}.
\end{eqnarray}
In such model, implying the properties of isotropic opticaly non-active media \cite{book:Kielich, book:Boyd}, the  third order nonlinear part of polarization supposed to have the form:
\begin{equation}\label{eq_nonlinear_third_rozp}
(P_{NL})_{y} = 4\pi\Big(\chi^{(3)}_{yyyy}E^{3}_{y}\Big).
\end{equation}
These results will certainly simplify the further derivation procedure in the part of nonlinearity account. However up to this point we are unable to tell whether the unidirectional approximation would become more or less robust for description of pulse propagation. To establish a point of reference we provide steps which lead to SPE equation through application of projection operators.\\

\subsection{On projection method.}

As a first step to illustrate the method \cite{book:Leble0} we define projection operators for a simple case of linear isotropic dielectric media where $D_{y} = \epsilon E_{y}$ and $\epsilon = 1 + 4\pi\chi^{1}$. In this case we can rewrite Eq. (\ref{eq_new_Maxwell's})
as matrix equation of a type 
$$\psi_t=L\psi,$$ 
where the field vector
\begin{equation}
\psi=\left(
          \begin{array}{c}
               E_z \\
               B_y \\
          \end{array}
          \right)\textrm{ and } L=\left(
          \begin{array}{cc}
               0 & \frac{c}{\epsilon}\partial_{x}\\
               c\partial_{x} & 0\\
          \end{array}
          \right),
\end{equation}
enter the matrix operator equation as
\begin{equation}\label{eq:matr_diel2}
\left(\begin{array}{c}
    (E_z)_t \\
    (B_y)_t \\
\end{array}\right) = \left(
          \begin{array}{cc}
            0 & \frac{c}{\epsilon}\partial_{x}\\
            c\partial_{x} & 0\\
          \end{array}
          \right)\left(\begin{array}{c}
            E_z \\
            B_y \\
          \end{array}
          \right) = \left(\begin{array}{c}
 			 \frac{c}{\epsilon}(B_y)_x\\
  				c(E_z)_x \\
			\end{array}\right).
\end{equation}

Consider now a Cauchy problem for the system Eq. (\ref{eq:matr_diel2}). 
Applying Fourier transformation on $x$ and hence rewriting this equation in frequency domain, one arrives at
ordinary equations system
\begin{equation}\label{eq:matr_diel2h}
\left(\begin{array}{c}
    (\omega\hat{E}_{z}(k,\omega)) \\
    (\omega\hat{B}_{y}(k,\omega)) \\
\end{array}\right) = \left(\begin{array}{c}
 			\frac{ck}{\epsilon}(\hat{B}_{y}(k,\omega))\\
  			kc(\hat{E}_{z}(k,\omega)) \\
	\end{array}\right).
\end{equation}
Let us next search for such matrix $P_{i}$ that $P_{1} \Psi = \Psi_{1}$ and $P_{2} \Psi= \Psi_{2}$ to be eigenvectors of the evolution matrix in Eq. (\ref{eq:matr_diel2h}). Moreover, the standard properties of orthogonal projecting operators
\begin{equation}\label{eq_matr_proj}
P_{i}*P_{j} = 0, \quad P^{2}_{i} = P_{i}, \quad \sum_{i}P_{i} = 1
\end{equation}
are implied. For Eq. (\ref{eq:matr_diel2h}) $P_{i}$ get the form
\begin{equation}\label{eq_matr_proj_diel}
P_{1} = \frac{1}{2}\left(\begin{array}{cc}
1 & \frac{1}{\sqrt{\epsilon}}\\
\sqrt{\epsilon} & 1\\
\end{array}\right), \quad P_{2} = \frac{1}{2}\left(\begin{array}{cc}
1 & -\frac{1}{\sqrt{\epsilon}}\\
-\sqrt{\epsilon} & 1\\
\end{array}\right).
\end{equation}
Performing the inverse Fourier transform yields the x-representation of the operators \cite{book:Leble0}.In this simplest case we consider the matrix elements of projecting operators do not depend on $k$, hence its x-representation coinside with the k-representation Eq. (\ref{eq_matr_proj_diel}).

With  the projection operators we can introduce new variables $\Lambda$ and $\Pi$
\begin{eqnarray}\label{eq_new_varible}
P_{1}\psi &=& \left(\begin{array}{c}
\frac{1}{2}E_{z} + \frac{1}{2\sqrt{\epsilon}}B_{y}\\
\frac{1}{2}\sqrt{\epsilon}E_{z} + \frac{1}{2}B_{y}\\
\end{array}\right) = \left(\begin{array}{c}
				\Lambda\\
				\sqrt{\epsilon}\Lambda\\
\end{array}\right),\nonumber\\
P_{2}\psi &=& \left(\begin{array}{c}
	\frac{1}{2}E_{z} - \frac{1}{2\sqrt{\epsilon}}B_{y}\\
	- \frac{1}{2}\sqrt{\epsilon}E_{z} + \frac{1}{2}B_{y}\\
\end{array}\right) = \left(\begin{array}{c}
				\Pi\\
				-\sqrt{\epsilon}\Pi\\
\end{array}\right),
\end{eqnarray}
which correspond to left and right direction of wave  propagation \cite{book:Leble0}. Moreover comparing new variables $\Lambda$ and $\Pi$ to variables presented by Kinsler et. al. \cite{Kins2} ours have similar form to their ones. The differences are caused by the fact that we have taken into account linear interaction between pulse and media. However our form of Eq. (\ref{eq_new_varible}) is exactly determined by dispersion relation $\omega = \frac{ck}{\sqrt{\epsilon}}$ from Eq. (\ref{eq:matr_diel2h}). What is more this allows us to present in simple way both electric and magnetic field.\cite{book:Leble}
\begin{equation}\label{eq:field_in_new_variable}
E_{z} = \Lambda + \Pi \quad B_{y} = \sqrt{\epsilon}(\Lambda - \Pi).
\end{equation} 
This correspondence Eq. (\ref{eq_new_varible},\ref{eq:field_in_new_variable}) is one-to-one local map and hence allows to determine initial conditions in the Cauchy problem for both left and right wave variables ($\Lambda,\Pi$).
It also gives a possibility to follow waves, extracting data in each time $t$.
The example we consider is simple, but contains all principle ingredients of the method \cite{book:Leble0}. More complicated example of the next section show what changes if the projection operator matrix elements depend on $k$. 

\subsection{Dispersion account: unidirectional waves subspaces.}

Within this formalism we reproduce calculations presented in \cite{SchWayne} going down to a unidirectional waves subspace. If next one assumes that medium of propagation can be modeled as made of free atoms interacting with the external electromagnetic field, one can present an expression on $\chi^{(1)}$ of the form \cite{book:Boyd}
\begin{equation}
\chi^{(1)} = c_{\chi}\sum_{n}|u_{n}|^{2}\left\{\frac{2\omega_{na}}{(\omega^{2}_{na}-\omega^{2}) +\gamma^{2}_{na}-2i\gamma_{na}\omega}\right\}.
\end{equation}
Then for silica fibers and for light in the visible to mid-infrared range there are three resonances of importance which occur at wavelengths of $\lambda = 0,068\dots\mu m$, $\lambda = 0,116\dots \mu m$ and $\lambda = 9,896\dots \mu m$. With those values we have approximated $\chi^{(1)}$ by
\begin{equation}\label{eq:Sch_chi0}
\chi^{(1)} = \frac{0.696\lambda^{2}}{\lambda^{2}-(0.0684)^{2}} + \frac{0.4079\lambda^{2}}{\lambda^{2}-(0.116)^{2}} +\frac{0.8974\lambda^{2}}{\lambda^{2}-(0.986)^{2}},
\end{equation}
where $\lambda$ is the wavelength expressed in $\mu m$. Following \cite{SchWayne} we focus on the propagation of light in the infrared range with $\lambda = 1600 - 3000nm$. In this range we can approximate
\begin{equation}\label{eq:Sch_chi1}
\chi^{(1)}(\lambda) \approx \chi^{(1)}_0 + \chi^{(1)}_{2}\lambda^{2}.
\end{equation}
If we plug Eq. (\ref{eq:Sch_chi1}) in $\epsilon$ and to
Eq. (\ref{eq_matr_proj_diel}) then the projection operators get the form
\begin{equation}\label{eq_matr_proj_diel_new_1}
P_{1} = \frac{1}{2}\left(\begin{array}{cc}
1 & \frac{1}{\sqrt{1 + 4\pi \chi^{(1)}_0 + 4\pi \chi^{(1)}_{2}\frac{{c}^{2}}{k^{2}}}}\\
\sqrt{1 + 4\pi \chi^{(1)}_0 + 4\pi \chi^{(1)}_{2}\frac{{c}^{2}}{k^{2}}} & 1\\
\end{array}\right),
\end{equation}
with the dependence on $k$ that arises from deispersion account. Similar form has the second operator 
\begin{equation}\label{eq_matr_proj_diel_new_2}
P_{2} = \frac{1}{2}\left(\begin{array}{cc}
1 & -\frac{1}{\sqrt{1 + 4\pi \chi^{(1)}_0 + 4\pi \chi^{(1)}_{2}\frac{{c}^{2}}{k^{2}}}}\\
-\sqrt{1 + 4\pi \chi^{(1)}_0 + 4\pi \chi^{(1)}_{2}\frac{{c}^{2}}{k^{2}}} & 1\\
\end{array}\right).
\end{equation}
To get the x-representation for the result we expand the coefficients in Taylor series and perform inverse Fourier transformation.
With new projection operators we rewrite relations Eq. (\ref{eq_new_varible}) in new conditions
\begin{eqnarray}\label{eq_new_varible_diel}
\Lambda &=& \frac{1}{2}\left(\sqrt{1+ 4\pi\chi^{(1)}_{0}}E_{z} + \frac{4\pi\chi^{(1)}_{2}c^{2}}{2\sqrt{1+ 4\pi\chi^{(1)}_{0}}}(i\partial x)^{-2}E_{z} + B_{y}\right),\\\nonumber
\Pi &=& \frac{1}{2}\left(-\sqrt{1+ 4\pi\chi^{(1)}_{0}}E_{z} - \frac{4\pi\chi^{(1)}_{2}c^{2}}{2\sqrt{1+ 4\pi\chi^{(1)}_{0}}}(i\partial x)^{-2}E_{z} + B_{y}\right).
\end{eqnarray}

\section{Nonlinearity account: general dynamics equations, SPE revisited}
\subsection{Nonlinear terms as perturbation}
If we account the  nonlinearity Eq. (\ref{eq_nonlinear_third_rozp}) then Eq. (\ref{eq:matr_diel2}) get the form
\begin{equation}\label{eq:matr_nonlin_evol}
\frac{\partial}{\partial t}\Psi - \mathbb{L}\Psi = \mathbb{N}(\Psi).
\end{equation}
Simplifying Eq. (\ref{eq_polariz}) to $ 4\pi\frac{\partial}{\partial t}\frac{1}{\epsilon}\chi^{(3)}E^{3}_{y}$ \cite{SchWayne, agrawal:book:NonFibOpt}
\begin{equation}\label{eq_matr_nonlin_diel}
\frac{\partial}{\partial t}\left(\begin{array}{c}
E_{y}\\
B_{z}\\
\end{array}\right) - \left(\begin{array}{cc}
			0 & \frac{c}{\epsilon}\partial_{x}\\
         c\partial_{x} & 0\\
\end{array}\right)\left(\begin{array}{c}
E_{y}\\
B_{z}\\
\end{array}\right) = 4\pi\frac{\partial}{\partial t}\left(\begin{array}{c}
\frac{1}{\epsilon}\chi^{(3)}E^{3}_{y}\\
0 \\
\end{array}\right).
\end{equation}
Applying $P_{1}$ Eq. (\ref{eq_matr_proj_diel}) on the LHS of Eq. (\ref{eq_matr_nonlin_diel}) and respect the projectors property $[P_{i},\left(\frac{\partial}{\partial t} - \mathbb{L}\right)]=0$, yields
\begin{equation}\label{eq:matr_nonlin_evol_proj_1}
\left(\frac{\partial}{\partial t} - \mathbb{L}\right)P_{i}\Psi = P_{i}\mathbb{N}(\Psi).
\end{equation}
Finally, present the result
\begin{equation}\label{eq:new_matr_nonlin_evol}
\left(\begin{array}{c}
\frac{\partial}{\partial t}\Lambda\\
\frac{\partial}{\partial t}\sqrt{\epsilon}\Lambda
\end{array}\right)-\left(\begin{array}{c}
\frac{c}{\epsilon}\frac{\partial }{\partial x}\sqrt{\epsilon}\Lambda\\
c\frac{\partial }{\partial x}\Lambda\\
\end{array}\right) = 4\pi\frac{\partial}{\partial t}\left(\begin{array}{c}
\frac{1}{\epsilon}\chi^{(3)}(\Lambda + \Pi)^{3}\\
\frac{1}{\sqrt{\epsilon}}\chi^{(3)}(\Lambda + \Pi)^{3} \\
\end{array}\right).
\end{equation}
Repeating our calculations from Eq. (\ref{eq:matr_nonlin_evol}) to Eq. (\ref{eq:new_matr_nonlin_evol}) with use of second projector operator Eq. (\ref{eq_matr_proj_diel_new_2}) than we will obtain a system of equations which describes interaction between two waves propagating in opposite directions. The system of equations have the form
\begin{equation}\label{eq:nonlin_two_ways}
\left\{\begin{array}{c}
\left(\begin{array}{c}
\frac{\partial}{\partial t}\Pi\\
-\frac{\partial}{\partial t}\sqrt{\epsilon}\Pi
\end{array}\right)-\left(\begin{array}{c}
\frac{c}{\epsilon}\frac{\partial }{\partial x}\sqrt{\epsilon}\Pi\\
-c\frac{\partial }{\partial x}\Pi\\
\end{array}\right) = 2\pi\frac{\partial}{\partial t}\left(\begin{array}{c}
\frac{1}{\epsilon}\chi^{(3)}(\Lambda + \Pi)^{3}\\
-\frac{1}{\sqrt{\epsilon}}\chi^{(3)}(\Lambda + \Pi)^{3}\\
\end{array}\right)\\
\left(\begin{array}{c}
\frac{\partial}{\partial t}\Lambda\\
\frac{\partial}{\partial t}\sqrt{\epsilon}\Lambda
\end{array}\right)-\left(\begin{array}{c}
\frac{c}{\epsilon}\frac{\partial }{\partial x}\sqrt{\epsilon}\Lambda\\
c\frac{\partial }{\partial x}\Lambda\\
\end{array}\right) = 2\pi\frac{\partial}{\partial t}\left(\begin{array}{c}
\frac{1}{\epsilon}\chi^{(3)}(\Lambda + \Pi)^{3}\\
\frac{1}{\sqrt{\epsilon}}\chi^{(3)}(\Lambda + \Pi)^{3} \\
\end{array}\right)
\end{array}\right. ,
\end{equation}
where $\epsilon = 1 + 4\pi \chi^{(1)}_0 + 4\pi \chi^{(1)}_{2}{c}^{2}(i\partial x)^{-2} $  is the operator in x-representation, which inverse and other functions we understand via corresponding Taylor series expansions. 
The couples of equations in Eq. (\ref{eq:nonlin_two_ways}) are in fact equivalent.

\subsection{The generalized Shafer-Wayne SPE}
Let us choose the first equation from the second system.
\begin{eqnarray}\label{eq:matrix_nonlin_one_way_1}
&&\left(1 + 4\pi \chi^{(1)}_0 + 4\pi \chi^{(1)}_{2}{c}^{2}(i\partial x)^{-2} \right)\frac{\partial}{\partial t}\Lambda - c\left(\sqrt{1+ 4\pi\chi^{(1)}_{0}} + \frac{4\pi\chi^{(1)}_{2}c^{2}}{2\sqrt{1+ 4\pi\chi^{(1)}_{0}}}(i\partial x)^{-2}\right)\frac{\partial }{\partial x}\Lambda = \nonumber\\
&& = 2\pi \frac{\partial}{\partial t}\chi^{(3)}(\Lambda + \Pi)^{3}.	
\end{eqnarray}
If we mean a pulse launched from a right end of a fiber we can consider the only direction of propagation. Hence we have chosen $\Lambda$ as a dominating left wave. Assume therefore that $\Pi =0$ and differentiate twice with respect to x produces the generalized SP equation
\begin{eqnarray}\label{eq:matrix_nonlin_one_way_3}
&&-\left(1 + 4\pi \chi^{(1)}_0\right)\frac{\partial^{3}}{\partial x^{2}\partial t}\Lambda + 4\pi\chi^{(1)}_{2}{c}^{2}\frac{\partial}{\partial t}\Lambda + c\left(\sqrt{1+ 4\pi\chi^{(1)}_{0}}\right)\frac{\partial^{3}}{\partial x^{3}}\Lambda - c\left(\frac{4\pi\chi^{(1)}_{2}c^{2}}{2\sqrt{1+ 4\pi\chi^{(1)}_{0}}}\right)\frac{\partial }{\partial x}\Lambda =\nonumber\\ &&-2\pi \frac{\partial^{3}}{\partial x^{2}\partial t}\chi^{(3)}(\Lambda)^{3}.
\end{eqnarray}
Following \cite{SchWayne} we can make an multiple scales ansatz
\begin{equation}\label{eq_ansatz_multiple}
\Lambda(x, t) = \left(\kappa A_{0}(\phi, x_{1},x_{2})+\kappa^{2} A_{1}(\phi, x_{1},x_{2})+\dots\right)
\end{equation}
where $\phi = \frac{t-x}{\kappa}$ and $x_{n} = \kappa^{n}x$. If we consider all terms $O(\kappa^{0})$ then we reach the final result
\begin{eqnarray}\label{eq:new_SPE}
&& -\left(3c\sqrt{1+ 4\pi\chi^{(1)}_{0}} - 2\big(1 + 4\pi \chi^{(1)}_0\big)\right) \frac{\partial^{2}}{\partial \phi\partial x_{1}}A_{0} - \left(4\pi\chi^{(1)}_{2}{c}^{2} - c\frac{4\pi\chi^{(1)}_{2}c^{2}}{2\sqrt{1+ 4\pi\chi^{(1)}_{0}}}\right)A_{0}\nonumber\\
&&=2\pi\chi^{(3)}\frac{\partial^{2}}{\partial\phi^{2}}(A_{0})^{3}.
\end{eqnarray}
At this point we obtain a second order differential equation which describes ultra short pulse launched in one direction. To fulfil this description we have to present initial-boundary conditions, that is immediate corollary from the projection fixed by Eq. (\ref{eq_new_varible_diel}) within the choice $\Pi=0$. It reads as the correlated action of magnetic and electric fields.
 
\section{Discussion and conclusions}

As we derive the system of equations Eq. (\ref{eq:nonlin_two_ways}) describing interaction between two waves propagating in two directions, we would like to remark that such interaction is weak in case of a long optical fiber exited from both ends.  However the significance of this phenomena is obvious: it give a new possibility investigate a nonlinearity and measure nonlinear constants.  The fundamental importance of the opposite waves interaction phenomena account is  quite clear in a case of optical resonators, when the effects  of interaction are cumulated. Going down to unidirectional case we acquire a SPE equation with modified coefficients within the demonstration of the general method \cite{book:Leble0}. One of the purposes of this publication is to attract attention of researches in optics and other fields to the projecting operators method. Some interesting features of the projection operator method applications to optical problems can be also seen in the paper of Kolesik et. al. \cite{Kolesik1}. In acoustics there are many interesting applications and important development with nonlinearity account in projecting operator recently published in Ref. \cite{Per} with references therein. 

We have also demonstrated how to obtain SPE with a projection operators method for unidirectional pulse propagation in isotropic, dispersionless media containing nonlinearity of the third order. The achieved results gives generalized SPE and correction to coefficients in it.


\begin{thebibliography}{99}
\bibitem{SchWayne} T. Sch\"{a}fer, G.E.Wayne, \textit{Propagation of ultra-short optical pulses in nonlinear media} (Elsevier Science, 2002).

\bibitem{ChungSchafer} Y. Chung, C.K.R.T. Jones, T. Sch\"{a}fer, and C. E. Wayne, \textit{Ultra-short pulses in linear and nonlinear media}. (Nonlinearity, 18:1351–1374, 2005).

\bibitem{Sakovich} S. Sakovich, \textit{Integrability of the Vector Short Pulse Equation}, (J. Phys. Soc. Jpn. 77, 2008).

\bibitem{SakovichSakovich} A. Sakovich, S. Sakovich, \textit{Solitary wave solutions of the short pulse equation}, (J. Phys. A: Math. Gen. 39, 2006).

\bibitem{Kins1} P. Kinsler, \textit{Unidirectional optical pulse propagation equation for materials with both electric and magnetic responses}, (Phys. Rev. A 81, 023808, 2010).

\bibitem{Kins4} P Kinsler, \textit{Transverse limits on the uni-directional pulse propagation approximation}, (arXiv:0810.5701).

\bibitem{Kins3} P. Kinsler, \textit{Limits of the uni-directional pulse propagation approximation}, (J. Opt. Soc. Am. B24, 2363-2368, 2007).

\bibitem{book:Leble0} S. Leble, \textit{Nonlinear waves in waveguides} (Springer, 1990).

\bibitem{book:Leble} S. Leble, \textit{Nonlinear Waves in Optical Waveguides and Soliton Theory Applications. Optical solitons, Theoretical and Experimental Challenges}, (pp 71-104, K. Porsezian, V.C. Kuriakose (Eds) Springer 2003).

\bibitem{Kins2} P. Kinsler, S. B. P. Radnor, G. H. C. New, \textit{Theory of directional pulse propagation}, (Phys. Rev. A72, 063807, 2005).

\bibitem{book:Kielich} S. Kielich, \textit{Nonlinear Molecular Optics} (PWN 1977).

\bibitem{book:Boyd} R.W. Boyd, \textit{Nonlinear Optics}. (Academic Press, Boston, 1992).

\bibitem{agrawal:book:NonFibOpt} G.P.Agrawal, \textit{Nonlinear fiber optics}, (Academic Press, 1997).

\bibitem{Kolesik1} M. Kolesik, J.V. Moloney, M. Mlejnek, \textit{Unidirectional Optical Pulse Propagation Equation}, (Phys. Rev. Lett. 89.283902, 2002)

\bibitem{Per}  A. Perelomova, \textit{Development of linear projecting in studies of non-linear flow. Acoustic heating induced by non-periodic sound},  (Physics Letters A 357,2006, 42-47).


\end{thebibliography}
\end{document}